\input harvmac
\input epsf
\newcount\figno
\figno=0
\def\fig#1#2#3{
\par\begingroup\parindent=0pt\leftskip=1cm\rightskip=1cm\parindent=0pt
\baselineskip=11pt
\global\advance\figno by 1
\midinsert
\epsfxsize=#3
\centerline{\epsfbox{#2}}
\vskip 12pt
{\bf Fig.\ \the\figno: } #1\par
\endinsert\endgroup\par
}
\def\figlabel#1{\xdef#1{\the\figno}}
\def\encadremath#1{\vbox{\hrule\hbox{\vrule\kern8pt\vbox{\kern8pt
\hbox{$\displaystyle #1$}\kern8pt}
\kern8pt\vrule}\hrule}}

\lref\rSen{A. Sen, "Descent Relations Among Bosonic D-branes," Int. J.
Mod. Phys. {\bf A14} (1999) 4061, hep-th/9902105.}
\lref\rOthers{A. Recknagel and V. Schomerus, "Boundary deformation theory
and moduli spaces of D-branes," Nucl. Phys. {\bf B545} (1999) 233, hep-th/9811237\semi
C.G. Callan, I.R. Klebanov, A.W. Ludwig and J.M. Maldacena, "Exact solution of
a boundary conformal field theory," Nucl. Phys. {\bf B422} (1994) 417, 
hep-th/9402113\semi
J. Polchinski and L. Thorlacius, "Free fermion representation of a boundary
conformal field theory," Phys. Rev. {\bf D50} (1994) 622, hep-th/9404008.}
\lref\rSntwo{A. Sen, "Stable non-BPS bound states of BPS D-branes," JHEP {\bf 9808},
(1998) 010, hep-th/9805019.}
\lref\rSnthr{A. Sen, "Tachyon condensation on the brane antibrane system," JHEP
{\bf 9808} (1998) 012, hep-th/9805170.}
\lref\rSnfr{A. Sen, "$SO(32)$ spinors of type I and other solitons on brane-antibrane
pait," JHEP {\bf 9809} (1998) 023, hep-th/9808141.}
\lref\rWitten{E. Witten, "D-branes and K-theory," JHEP {\bf 12} (1998) 019,
hep-th/9810188.}
\lref\rSnfv{A. Sen, "BPS D-branes on non-supersymmetric cycles," JHEP {\bf 12} (1998)
021, hep-th/9812031.}
\lref\rHrv{P. Horava, "Type IIA D-branes, K-theory and matrix theory," Adv. Theor.
Math. Phys. {\bf 2} (1999) 1373, hep-th/9812135.}
\lref\rone{A. Sen, "Universality of the Tachyon Potential," JHEP {\bf 9912} 027,
hep-th/9911116.}
\lref\rSZ{A. Sen and B. Zweibach, "Tachyon Condensation in String Field Theory,"
hep-th/9912249.}
\lref\rTay{W. Taylor, "D-brane effective field theory from string field theory,"
hep-th/0001201.}
\lref\rMT{N. Moeller and W. Taylor, "Level truncation and the tachyon in open
bosonic string field theory," hep-th/0002237.}
\lref\rHK{J.A. Harvey and P.Kraus, "D-branes as unstable lumps in bosonic open
string field theory," hep-th/0002117.}
\lref\rUs{R. de Mello Koch, A. Jevicki, M. Mihailescu and R. Tatar, "Lumps and
p-branes in open string field theory," hep-th/0003031.}
\lref\rBerk{N. Berkovits, "The tachyon potential in open Neveu-Schwarz String Field
Theory," hep-th/0001084.}
\lref\rBSZ{N. Berkovits, A. Sen and B. Zweibach, "Tachyon condensation in superstring
field theory," hep-th/0002211.}
\lref\rSR{P. de Smet and J. Raeymaekers, "Level four approximation to the tachyon
potential in superstring field theory," hep-th/0003220.}
\lref\rIN{A. Iqbal and A. Naqvi, "Tachyon condensation on a non-BPS D-brane,"
hep-th/0004015.}
\lref\rMSZ{N. Moeller, A. Sen and B. Zweibach, "D-branes as tachyon lumps in string 
field theory," hep-th/0005036.}
\lref\rHKLM{J.A. Harvey, P. Krauss, F. Larsen and E.J. Martinec, "D-branes and strings
as non-commutative solitons," hep-th/0005031.}
\lref\rWittn{E. Witten, "Noncommutative tachyons and string field theory," 
hep-th/0006071.}
\lref\rAnt{A. Bogojevic, A. Jevicki and G. Meng, "Quartic Interactions in
Superstring Field Theory," Brown preprint HET-672, 1988, available from KEK
preprint library.}
\lref\rKS{V.A. Kostelecky and S. Samuel, "The static tachyon potential in the
open bosonic string theory," Phys. Lett. {\bf B207} (1988) 169\semi
V.A. Kostelecky and R. Potting, "Expectation values, Lorentz Invariance and
CPT in the open bosonic string," Phys. Lett. {\bf B381} (1996) 89, hep-th/9605088.}
\lref\rRJ{R. de Mello Koch and J.P. Rodrigues, to appear.}
\lref\rGA{D. Gross and A. Jevicki, "Operator formulation of interacting string
field theory (I),(II)," Nucl. Phys. {\bf B283} (1987) 1, {\bf B287} (1987) 225\semi
E. Cremmer, A. Schwimmer and C. Thorn, "The vertex function in Wittens formulation
of string field theory," Phys. Lett. {\bf B179} (1986) 57\semi
S. Samuel, "The physical and ghost vertices in Witten's string field theory,"
Phys. Lett. {\bf B181} (1986) 255\semi
A. LeClair, M.E. Peskin and C.R. Preitschopf, "String field theory on the 
conformal plane. 1. Kinematical Principles," Nucl. Phys. {\bf B317} (1989) 411;
"String Field Theory on the conformal plane. 2. Generalized Gluing," Nucl. Phys.
{\bf B317} (1989) 464.}
\lref\rStrom{J.A. Harvey, P. Kraus, F. Larsen and E.J. Martinec, "D-branes
and strings as non-commutative solitons," hep-th/0005031\semi
K. Dasgupta, S. Mukhi and G. Rajesh, "Noncommutative tachyons," JHEP
{\bf 0006} (2000) 022, hep-th/0005006\semi
R. Gopakumar, S. Minwalla and A. Strominger, "Symmetry Restoration and
Tachyon Condensation in Open String Theory," hep-th/0007226.}
\lref\rMlr{N. Moeller, "Codimension two lump solutions in string field
theory and tachyonic theories," hep-th/0008101.}


\Title{ \vbox {\baselineskip 12pt\hbox{}
\hbox{}  \hbox{August 2000}}}
{\vbox {\centerline{Lumps in level truncated open string field theory}
}}

\smallskip
\centerline{Robert de Mello Koch and Jo\~ao P. Rodrigues}
\smallskip
\centerline{\it Department of Physics and Center for Nonlinear 
Studies,}
\centerline{\it University of the Witwatersrand,}
\centerline{\it Wits, 2050, South Africa}
\centerline{\tt robert,joao@physnet.phys.wits.ac.za}\bigskip

\noindent
The ratios of the masses of $D(p-d)$ branes to the masses of $Dp$
in open bosonic string field theory are computed within the modified 
level truncation approximation of Moeller, Sen and Zweibach. At the 
lowest non-trivial truncation requiring the addition of new primary 
states, we find evidence of rapid convergence to the expected result 
for $2\le d\le 6$ providing additional evidence for the consistency 
of this approximation.


\Date{}

\def\MAKEdalamSIGN#1#2{%
\setbox0=\hbox{$\mathsurround 0pt #1{#2}$}
\dimen0=\ht0 \advance\dimen0 -0.8pt
\hbox{\vrule\vbox to\ht0{\hrule width\dimen0 \vfil\hrule}\vrule}}


There is growing evidence that open string field theory provides 
a direct approach to study string theory tachyons. This recent 
progress has been possible as a result of Sen's conjecture that 
there is an extremum of the tachyon potential at which the total 
negative potential energy exactly cancels the tension of the 
$D$-brane\rSen\ and that lump solutions are identified with lower 
dimensional branes\refs{\rSen,\rOthers}. These conjectures have 
been extended for tachyons living on coincident $D$-brane 
anti-$D$-brane pairs and for tachyons on the non-BPS $D$-branes of 
type IIA or IIB superstring 
theories\refs{\rSntwo,\rSnthr,\rSnfr,\rWitten,\rSnfv,\rHrv}. 
These conjectures provide 
precise predictions which can be used to test and develop 
approximation techniques in open string field theory. The
results obtained thus far are 
impressive\refs{\rSZ,\rTay,\rMT,\rHK,\rUs,\rBerk,\rBSZ,\rSR,\rIN,\rMSZ,\rHKLM,\rWittn}. 
In particular, the level 
truncation approximation has proved to be powerful in this context.
The original argument for level truncation appeared in the unpublished 
work\rAnt\ and was subsequently used by Samuel and Kostelecky\rKS\ 
to study the vacuum structure of string field theory. 
In this article, we are interested in a variant of the level truncation 
scheme, introduced by Moeller, Sen and Zweibach (MSZ)\rMSZ. 
Within this scheme, MSZ were able to 
compute the ratio of the mass of a $D(p-1)$-brane to a $Dp$-brane to an 
impressive accuracy of about 1\%!\foot{By turning on a large $B$ field the 
description of tachyon condensation can be drastically simplified. In this
limit, the tension computed from tachyonic solitons exactly agree with the
expected $D$-brane tensions\rStrom.} 

This ratio was computed in \rHK\ in the field theory limit and good
agreement with the expected result was obtained. Indeed, the field
theory lump reproduces 78\% of the expected $D(p-1)$ brane tension.
Including the stringy corrections from the momentum dependence of the
interaction terms does not significantly change the tension of the lump.
This is not the case for the ratios of the masses of $D(p-d)$ branes to
the masses of $Dp$ branes for $d>1$. As $d$ is increased the field
theory predictions get worse and for $d$ large enough there are no
lump solutions. When the stringy corrections are included, lump
solutions can be found for any value of $d$\rUs. However, initial studies
of these ratios show that the leading order in the level truncation
approximation over estimates this ratio for $d>4$, with the results
becoming increasingly worse as $d$ is increased. As a further interesting
check of the modified level truncation scheme, one could compute
the ratios of the masses of $D(p-d)$ branes to the mass of a $Dp$
brane. These ratios are more difficult to reproduce and obtaining
them within the modified level truncation approximation will give
important insight into how fast the approximation converges. This
is the question that is studied in this article.

The modified level truncation approximation starts by assigning to a 
state $|\Phi_i\rangle$, with number eigenvalue $N_i$, the level

\eqn\Nlevel
{l(\Phi_{i,n})\equiv {\vec{n}\cdot\vec{n}\over R^2}+N_i-N_0,}

\noindent
where $N_0$ is the number eigenvalue for the zero momentum tachyon\rMSZ. The
level $(M,N)$ approximation to the action is then defined by keeping only
fields with level $\le M$ and terms in the action for which the sum of 
levels is $\le N$. We assume that the background spacetime is the product
of a $(d+1)+1$ dimensional flat spacetime labelled by the spacelike
coordinates $(x^1,...,x^d,y)$ and a timelike coordinate $x^0$, and a 
Euclidean manifold ${\cal M}$ described by a conformal field theory of 
central charge $26-d-2$. The spatial direction $y$ is non-compact; the 
$x^i$, $i=1,...,d$ parametrize a torus $T^d$: $x^i\sim x^i+R.$ As in \rone\
by studying the motion of the brane along this non-compact direction, we 
can compute its tension. For an open string ending on the $D$-brane, we 
put Neumann boundary conditions on the fields $(X^1,...,X^d)$ and $X^0$, 
and Dirichlet boundary conditions on the field $Y$ and on the fields 
associated with the coordinates of ${\cal M}$. These boundary conditions 
are correct for a $Dd$-brane wrapped on the $T^d$. Because the $D$-brane
has a finite volume, it will have a finite mass.

The dynamics of the open strings with ends on this $D$-brane is described
by the direct sum of the conformal field theories associated with the fields
$X^i$, $Y$, $X^0$ and the manifold ${\cal M}$. Following \rone\ we will work on 
a subspace of the full string field theory Fock space. Towards this end,
we denote the conformal field of the $X^i$ by CFT($T^d$) and the conformal
field theory of the fields $Y$, $X^0$ and of the manifold ${\cal M}$ by
CFT$'=$CFT($Y$)$\oplus$CFT($X^0$)$\oplus$CFT(${\cal M}$). The Virasoro
generators of the system are given by $L_n=L_n^{ghost}+L_n^{T^d}+L_n'$, where
$L_n^{ghost}$ are the Virasoro generators of the ghost system, $L_n^{T^d}$ are 
the Virasoro generators of CFT($T^d$) and $L_n'$ are the Virasoro generators of
CFT$'$. The subspace of the full string Fock space that we will focus on is
most easily characterized by grouping states into Verma modules\rone. Each Verma
module can be labelled by a primary state. The Verma module contains this 
primary state together will all states obtained by acting on this primary with 
the Virasoro generators $L_{-n}^{T^d},L_{-n}',c_1,c_{-n}$ and $b_{-n}$ with
$n\ge 0$. Null states and their descendents should be removed. The truncation 
can now be described by specifying which primary states we consider. We consider
primary states (with arbitrary momentum on the $T^d$) of CFT($T^d$) which are 
even under $X\to -X$ and are trivial CFT$'$ primaries. We will also restrict 
to states of even twist. For more details, 
the reader should consult\refs{\rone\rMSZ}. Working on this subspace 
was shown in\refs{\rone,\rMSZ} to be 
a consistent truncation of the full open string field theory. 

If we work to a given level, then we need only consider the
Verma modules built on a finite number of primary fields. In
our subspace, the zero momentum tachyon mode is the only
primary that needs to be considered at level $0$; there are
$d-1$ zero-momentum primaries to be considered at level $2$
\rRJ. In the analysis of \rMSZ, the case $d=1$ was considered,
so no new zero-momentum primaries had to be added until level 4.
Since we work only to level $(2,4)$ we need not consider any 
other zero-momentum primaries. Thus, the states that we consider are

\eqn\Tach
{|T_{\vec{n}}\rangle= c_1\cos
\Big({\vec{n}\cdot\vec{X}(0)\over R}\Big)|0\rangle }

\eqn\U
{|U_{\vec{n}}\rangle= c_{-1}\cos
\Big({\vec{n}\cdot\vec{X}(0)\over R}\Big)|0\rangle }

\eqn\V
{|V_{\vec{n}}\rangle= c_1 L_{-2}^{T^d}\cos
\Big({\vec{n}\cdot\vec{X}(0)\over R}\Big)|0\rangle ,}

\eqn\W
{|W_{\vec{n}}\rangle= c_1 L_{-2}^{\prime}\cos
\Big({\vec{n}\cdot\vec{X}(0)\over R}\Big)|0\rangle ,}

\eqn\Z
{|Z_{\vec{n}}\rangle= c_1 L_{-1}^{T^d}L_{-1}^{T^d}\cos
\Big({\vec{n}\cdot\vec{X}(0)\over R}\Big)|0\rangle ,}

\eqn\S
{|S^i_{\vec{n}}\rangle= c_1 (\alpha_{-1}^1\alpha_{-1}^1
-\alpha_{-1}^{i+1}\alpha_{-1}^{i+1}-\alpha_{0}^1\alpha_{-2}^1
+\alpha_{0}^{i+1}\alpha_{-2}^{i+1})\cos
\Big({\vec{n}\cdot\vec{X}(0)\over R}\Big)|0\rangle .}

\noindent
The states $|S^i_{\vec{0}}\rangle$ are the new zero momentum primaries
that need to be added at level $2$.
We should not include the zero momentum $Z_{\vec{0}}$ mode, since it
corresponds to the descendent of a null state. The index $i$
runs from $i=1,...,d-1$. The conditions that the states 
$|S_{\vec{n}}^i\rangle$ are primary implies the following 
restriction on $\vec{n}$

\eqn\Restr
{n^1=\pm n^{i+1}.}

\noindent
The string field is expanded in terms of these states as follows

\eqn\StrFld
{|\Phi\rangle=\sum_{\vec{n}}\Big(t_{\vec{n}}|T_{\vec{n}}\rangle+
u_{\vec{n}}|U_{\vec{n}}\rangle+v_{\vec{n}}|V_{\vec{n}}\rangle+
w_{\vec{n}}|W_{\vec{n}}\rangle+z_{\vec{n}}|Z_{\vec{n}}\rangle
+s_{\vec{n}}^i|S_{\vec{n}}^i\rangle\Big).}

\noindent
To ensure that we have the $X^i\to -X^i$ symmetry we have to
put restrictions on the coefficients appearing in this
expansion. For example, in the case that $d=2$, states
carrying momentum $\vec{n}=(1,1)$ are related to states carrying
momentum $\vec{n}=(1,-1)$ by $X^2\to -X^2$. Thus, the coefficients
corresponding to these states in \StrFld\ should be identified.
If we were treating
this problem in the $R\to\infty$ limit, we'd look for solutions
which are rotationally invariant. However, for finite $R$, all
that survives of the rotational invariance is a discrete subset
of rotations which permute the different $X^i$. We will
compute the potential below assuming this symmetry. Thus for
example, in the case $d=2$, we will not distinguish between
states carrying momentum $\vec{n}=(0,1)$ and states carrying
momentum $\vec{n}=(1,0)$. In addition, we will not distinguish
between $s_0^i$ for diferent values of $i$.

The string field theory action is given by

\eqn\Action
{S=\langle\Phi|Q|\Phi\rangle+{g\over 3}\langle V_3||\Phi\rangle 
|\Phi\rangle|\Phi\rangle}

\noindent
with $Q$ the first quantized BRST operator and $|V_3\rangle$ the 
three string interaction vertex.
To compute the potential, we have used the operator representation of
the vertex as an object in the three string (dual) Fock space\rGA.
Evaluation of the above two terms in the action then requires only
algebraic manipulations. The details of this computation will
appear elsewhere\rRJ. The result, up to level $(2,4)$ is (with
$R=\sqrt{3}$)

\eqn\one
{V(0,0)=-{1\over 2}t_0^2 +{1\over 3}K^3 t_0^3 .}

\eqn\two
{V\Big({1\over 3},{2\over 3}\Big)=V(0,0)-{d\over 4}\Big(1-{1\over R^2}\Big)t_1^2+
{d\over 2}K^{3-{2\over R^2}}t_0 t_1^2 .}

\eqn\three
{V\Big({2\over 3},{4\over 3}\Big)=V\Big({1\over 3},{2\over 3}\Big)+
{d(d-1)\over 2}K^{3-{4\over R^2}}t_2^2 t_0-{d(d-1)\over 4}
\Big(1-{2\over R^2}\Big)t_2^2+{d(d-1)\over 2}K^{3-{4\over R^2}}
t_1^2 t_2.}

\eqn\four
{\eqalign{V(1,2)=V\Big({2\over 3},{4\over 3}\Big)+d(d-1)(d-2)\Big[&{1\over 3}
K^{3-{6\over R^2}}t_3^2 t_0-{1\over 6}t_3^2\Big(1-{3\over R^2}\Big)\cr
&+K^{3-{6\over R^2}}t_1 t_2 t_3
+{1\over 3}K^{3-{6\over R^2}}t_2^3\Big].}}

\eqn\five
{\eqalign{V\Big({4\over 3},{8\over 3}\Big)&=V(1,2)+d(d-1)(d-2)(d-3)\Big[
{1\over 6}\tilde{t}_4^2 t_0K^{3-{8\over R^2}}-{1\over 12}\tilde{t}_4^2
\Big(1-{4\over R^2}\Big)\cr
&+{1\over 2}\tilde{t}_4 t_2^2 K^{3-{8\over R^2}}
+{2\over 3}t_1 t_3 \tilde{t}_4 K^{3-{8\over R^2}}
+t_3^2 t_2K^{3-{8\over R^2}}\Big]
+d\Big[{1\over 2}\bar{t}_4^2 t_0 K^{3-{8\over R^2}}\cr
&-{\bar{t}_4^2\over 4}\Big(1-{4\over R^2}\Big)
+{d-1\over 2}
t_2^2\bar{t}_4K^{3-{8\over R^2}}+{1\over 4}t_1^2\bar{t}_4
K^{3-{6\over R^2}}\Big].}}

\eqn\six
{\eqalign{V\Big({5\over 3},{10\over 3}\Big)&=V\Big({4\over 3},{8\over 3}\Big)+
d(d-1)\Big(t_0\bar{t}_5^2 K^{3-{10\over R^2}}-{1\over 2}\bar{t}_5^2
\Big[1-{5\over R^2}\Big]+t_1 \bar{t}_4\bar{t}_5 K^{3-{10\over R^2}}\cr
&+2(d-2)t_2 t_3\bar{t}_5K^{3-{10\over R^2}}
+{1\over 2}(d-2)t_3^2 \bar{t}_4
K^{3-{10\over R^2}}\Big)+d(d-1)(d-2)(d-3)\times\cr
&\times(d-4)\Big({1\over 15}t_0\tilde{t}_5^2K^{3-{10\over R^2}}-{1\over 30}
\tilde{t}_5^2\Big[1-{5\over R^2}\Big]+{1\over 3}t_1 \tilde{t}_4\tilde{t}_5
K^{3-{10\over R^2}}+{2\over 3}t_2 t_3\tilde{t}_5K^{3-{10\over R^2}}\cr
&+{2\over 3}t_2 \tilde{t}_4^2K^{3-{10\over R^2}}
+t_3^2 \tilde{t}_4 K^{3-{10\over R^2}}\Big)+d(d-1)t_1 t_2 \bar{t}_5 
K^{3-{8\over R^2}},}}

\eqn\seven
{\eqalign{V(2,&4)=V\Big({5\over 3},{10\over 3}\Big)+d(d-1)(d-2)\Big(t_0\bar{t}_6^2
K^{3-{12\over R^2}}+2t_1\bar{t}_5\bar{t}_6K^{3-{12\over R^2}}+t_2\bar{t}_5^2
K^{3-{12\over R^2}}\cr
&+t_1 t_3\bar{t}_6K^{3-{10\over R^2}}+
{K^{1-{6\over R^2}}\over 3}\Big[{-5d\over 32}+{3\over R^2}\Big]t_3^2 v_0
+{11\over 48}K^{1-{6\over R^2}}u_0 t_3^2-{5(26-d)\over 96}
K^{1-{6\over R^2}} w_0 t_3^2\cr
&-{1\over 2}\bar{t}_6^2\Big[1-{6\over R^2}\Big]+t_2\bar{t}_4\bar{t}_6
K^{3-{12\over R^2}}\Big)+d(d-1)(d-2)(d-3)(d-4)(d-5)K^{3-{12\over R^2}}\times\cr
&\Big({1\over 45}t_0\tilde{t}_6^2+{2\over 15}t_1\tilde{t}_5\tilde{t}_6
+{1\over 3}t_2\tilde{t}_4\tilde{t}_6+{1\over 3}t_2\tilde{t}_5^2
+{2\over 9}t_3^2\tilde{t}_6+{4\over 3}t_3\tilde{t}_4
\tilde{t}_5+{2\over 3}\tilde{t}_4^3\Big)+d(d-1)\times\cr
&(d-2)(d-3)K^{3-{12\over R^2}}\Big(2t_3^2\bar{t}_6
+2t_3\tilde{t}_4\bar{t}_5+{1\over 3}\tilde{t}_4^2\bar{t}_4\Big)
+{19 K\over 144}u_0^2 t_0-{5d\over 32}Kt_0^2 v_0\cr
&-d(d-1)(d-2)(d-3)(d-4)(d-5){\tilde{t}_6^2\over 90}\Big(1-{6\over R^2}\Big)
+d(d-1)\Big({1\over 4}s_0^2\cr
&+{1\over 2}K^{1-{4\over R^2}}\Big[-{5d\over 32}+{2\over R^2}\Big]
t_2^2 v_0+{11\over 32}K^{1-{4\over R^2}}u_0 t_2^2
-{5(26-d)\over 64}K^{1-{4\over R^2}}w_0t_2^2+
{32\over 27}Ks_0^2 t_0\Big)\cr
&+{K\over 432}\Big[{25 d^2\over 4}+128d\Big]t_0v_0^2+
{d\over 2}K^{1-{2\over R^2}}\Big[-{5d\over 32}+{1\over R^2}\Big]t_1^2 v_0
+{11\over 32}dK^{1-{2\over R^2}}u_0 t_1^2\cr
&+{11\over 16}Kt_0^2 u_0-K{5(26-d)\over 32}t_0^2 w_0+{K\over 432}
\Big[{25(26-d)^2\over 4}+128(26-d)\Big]t_0w_0^2+{d\over 4}v_0^2\cr
&-{5(26-d)\over 64}dK^{1-{2\over R^2}}w_0 t_1^2-{55d\over 432}Kt_0 u_0 v_0
-{55(26-d)\over 432}Kt_0 u_0 w_0-{u_0^2\over 2}+{26-d\over 4}w_0^2\cr
&+{25d(26-d)\over 864}Kt_0 w_0 v_0+{1\over 2}d(d-1)t_2\bar{t}_5^2 
K^{3-{12\over R^2}}.}}

\noindent
In this potential we have labelled the modes by their
$\vec{n}\cdot\vec{n}$ eigenvalue. This labelling is not unique
if $d>3$ and $\vec{n}\cdot\vec{n}\ge 4$. For example, when
$d=4$ and $\vec{n}\cdot\vec{n}= 4$ we can build the $\vec{n}$ as $\vec{n}=(1,1,1,1)$
or $\vec{n}=(2,0,0,0).$ We distinguish between these two vectors by
using a tilde $(\tilde{t})$ to denote vectors whose only entries are
$1$s and by using a bar $(\bar{t})$ to denote vectors whose entries 
include a $2$. This labeling is unique to the level considered here. 
We now seek lump solutions for the tachyon condensate
which minimize this potential.

The lump solutions are local minima of the potential. A
direct minimization of the potential usually yields the
tranlationally invariant vacuum which is the unique
global minimum. To obtain the lump solutions, we found
it simplest to work with the equations of motion. The
numerical results obtained in this way were stable, both
as $d$ is increased and as the level is increased. For
example, the following table summarizes the values of the
fields at the minimum corresponding to the $D0$ lump,
with $R=\sqrt{3}$ and $d=2$

$$ \vbox{\offinterlineskip \hrule
\halign{&\vrule#&\strut\ \  \hfil#\ \ \cr
height2pt&\omit&&\omit&&\omit&&\omit&&\omit&&\omit&&\omit&\cr
\noalign{\hrule}
&~~~~ &&$({1\over 3},{2\over 3})$&&$({2\over 3},{4\over 3})$
&&$({4\over 3},{8\over 3})$&&$({5\over 3},{10\over 3})$&&$(2,4)$&\cr
\noalign{\hrule}
height2pt&\omit&&\omit&&\omit&&\omit&&\omit&&\omit&&\omit&\cr
\noalign{\hrule} 
height2pt&\omit&&\omit&&\omit&&\omit&&\omit&&\omit&&\omit&\cr
&$t_0$&& 0.1810 && 0.2863 && 0.3021 && 0.3186 && 0.3806&\cr 
\noalign{\hrule}
height2pt&\omit&&\omit&&\omit&&\omit&&\omit&&\omit&&\omit&\cr
&$t_1$&& 0.2435 && 0.2115 && 0.2025 && 0.1915 && 0.2127&\cr 
\noalign{\hrule}
height2pt&\omit&&\omit&&\omit&&\omit&&\omit&&\omit&&\omit&\cr
&$t_2$&&  && -0.1253 && -0.1256 && -0.1257 && -0.1432&\cr 
\noalign{\hrule}
height2pt&\omit&&\omit&&\omit&&\omit&&\omit&&\omit&&\omit&\cr
&$\bar{t}_4$&&  &&  && -0.0442 && -0.0495 && -0.0563&\cr 
\noalign{\hrule}
height2pt&\omit&&\omit&&\omit&&\omit&&\omit&&\omit&&\omit&\cr
&$\bar{t}_5$&&  &&  &&  && 0.0280 && 0.0353&\cr 
\noalign{\hrule}
height2pt&\omit&&\omit&&\omit&&\omit&&\omit&&\omit&&\omit&\cr
&$u_0$&&  &&  &&  &&  && 0.1268&\cr 
\noalign{\hrule}
height2pt&\omit&&\omit&&\omit&&\omit&&\omit&&\omit&&\omit&\cr
&$v_0$&&  &&  &&  &&  && 0.0234&\cr 
\noalign{\hrule}
height2pt&\omit&&\omit&&\omit&&\omit&&\omit&&\omit&&\omit&\cr
&$w_0$&&  &&  &&  &&  && 0.0417&\cr 
\noalign{\hrule}
height2pt&\omit&&\omit&&\omit&&\omit&&\omit&&\omit&&\omit&\cr
&$s_0^i$&&  &&  &&  &&  && 0.00&\cr 
\noalign{\hrule}
height2pt&\omit&&\omit&&\omit&&\omit&&\omit&&\omit&&\omit&\cr
\noalign{\hrule}}}$$

\noindent
The fact that $s_0^i$ is zero is a consequence of the fact that it enters the
action quadratically. Generically, at this level, the potential would include 
terms linear in $s_0^i$ which would couple $s_0^i$ to the $t_1$, $t_2$ and $t_3$
tachyon modes. However, due the symmetry of the tachyon solution these couplings
vanish once summed over momenta. 

We have used these minima to compute the ratio of the mass of a $D(p-d)$-brane
to the mass of a $Dp$ brane. The quantity that we measure is

$$r={E_{lump}\over T_0}={(2\pi R)^p\over 2^{p-1}\pi^{p-2}}\Big(
V_{lump}-V_{vac}\Big),$$

\noindent
where $E_{lump}$ in the energy of the lump solution, $T_0$ is the mass of
a zero brane, $V_{lump}$ is the value of the potential evaluated at the lump 
solution and $V_{vac}$ is the value of the potential evaluated at the global minimum 
corresponding to the translationally invariant tachyon condensate. The predicted 
value for this ratio is 1. We have computed $V_{lump}$ and $V_{vac}$ to the same 
order in the modified level truncation expansion. In \rMSZ\ this parameter was called 
$r^{(2)}$. The values of $r$ obtained in this study are

$$ \vbox{\offinterlineskip \hrule
\halign{&\vrule#&\strut\ \  \hfil#\ \ \cr
height2pt&\omit&&\omit&&\omit&&\omit&&\omit&&\omit&&\omit&&\omit&\cr
\noalign{\hrule}
height2pt&\omit&&\omit&&\omit&&\omit&&\omit&&\omit&&\omit&\cr
&~~~~ &&$({1\over 3},{2\over 3})$&&$({2\over 3},{4\over 3})$
&&$(1,2)$&&$({4\over 3},{8\over 3})$&&$({5\over 3},{10\over 3})$
&&$(2,4)$&\cr
\noalign{\hrule}
height2pt&\omit&&\omit&&\omit&&\omit&&\omit&&\omit&&\omit&&\omit&\cr
\noalign{\hrule}
height2pt&\omit&&\omit&&\omit&&\omit&&\omit&&\omit&&\omit&\cr
&$d=2$&& 1.3402 && 0.8992 &&  && 0.8377 && 0.7772 && 1.1303&\cr
\noalign{\hrule}
height2pt&\omit&&\omit&&\omit&&\omit&&\omit&&\omit&&\omit&\cr
&$d=3$&& 2.3213 && 1.4237 && 1.0690 && 1.0278 && 0.9313 && 1.3277 &\cr
\noalign{\hrule}
height2pt&\omit&&\omit&&\omit&&\omit&&\omit&&\omit&&\omit&\cr
&$d=4$&& 4.0206 && 2.3658 && 1.5659 && 1.2712 && 1.1754 && 1.6384 &\cr
\noalign{\hrule}
height2pt&\omit&&\omit&&\omit&&\omit&&\omit&&\omit&&\omit&\cr
&$d=5$&& 6.963 && 4.009 && 2.472 && 1.7782 && 1.5036 && 2.0901 &\cr
\noalign{\hrule}
height2pt&\omit&&\omit&&\omit&&\omit&&\omit&&\omit&&\omit&\cr
&$d=6$&& 12.06 && 6.8225 && 4.0391 && 2.6862 && 2.0618 && 2.6641 &\cr
\noalign{\hrule}
height2pt&\omit&&\omit&&\omit&&\omit&&\omit&&\omit&&\omit&\cr
\noalign{\hrule}}}$$

\noindent
The decrease in the value of $r$ as higher tachyon modes are added
and the sharp increase in the tension at level (2,4) qualitatively match
the results obtained in \rMSZ.

For $d=2$ we have ploted the tachyon profile of the $D0$ brane lump
solution at levels $(2,4)$ and $(5/3,10/3)$

\fig{A contour plot of the tachyon lump profile at level $(2,4)$
and at radius $R=\sqrt{3}$. We have reversed the signs of $t_1$ and $\bar{t}_5$,
which shifts the lump's center from $x=y=\pi R$ to $x=y=0$.}{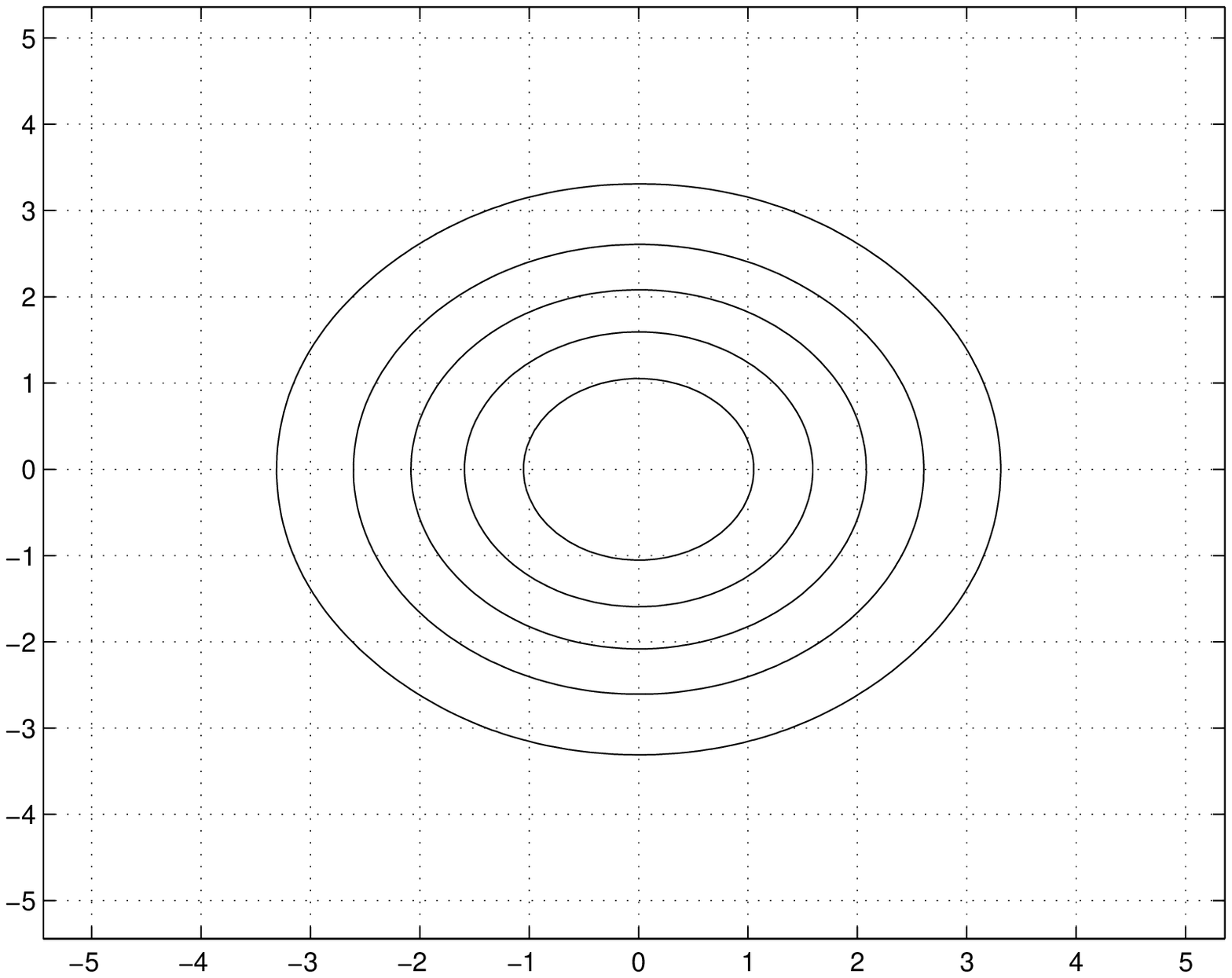}{2.8truein}

\vfill\eject

\fig{A profile plot of the tachyon lump at level $({6\over 3},{12\over 3})$
and at radius $R=\sqrt{3}$.}{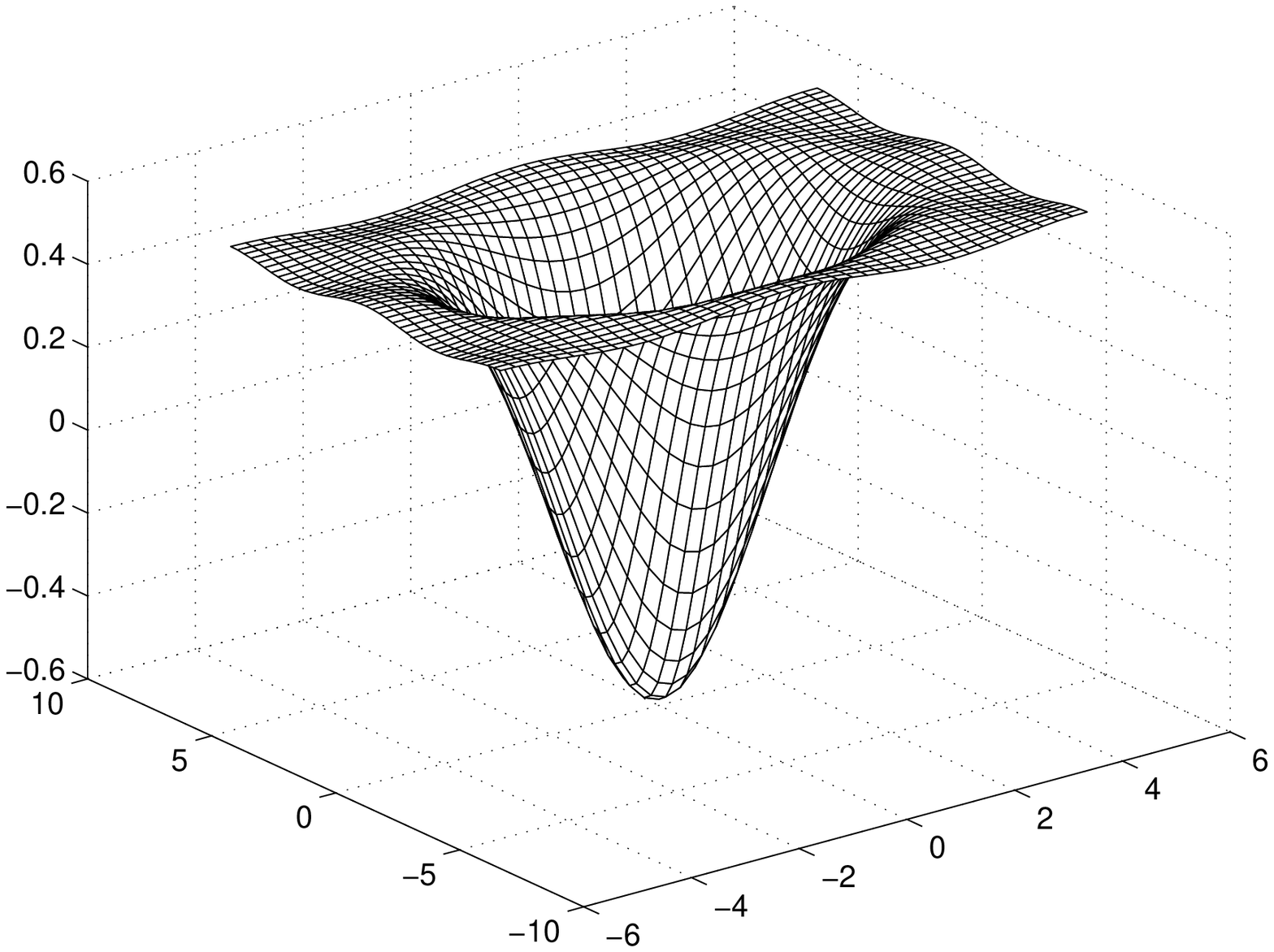}{2.8truein}

Our results indicate the the modified level truncation proposed
by MSZ provides a powerful tool for the study of tachyon
condensation in bosonic open string field theory. It is interesting
to extend these results to higher level, to further test the accuracy
and concergence of the modified level truncation approximation.
It is possible to continue to level $(11/3,22/3)$ with $R=\sqrt{3}$
before we need to add any new zero-momentum primaries. This study is
in progress. 

\noindent
{\it Note Added:} An independent study of two dimensional solitons in open
bosonic string theory has recently appeared in \rMlr. The results of \rMlr\ 
agree with the results we obtained here for $d=2$.

\noindent
{\it Acknowledgements:} We would like to thank Nicolas Moeller for helpful
correspondence and detailed comparison. This was crucial for correcting a
coefficient in equation (17). This research was supported in part by the NRF under
Grant No. GUN-2034479. RdMK also thanks the University of the Witwatersrand for
a Friedel Schellschop award. 

\listrefs
\vfill\eject
\bye